\documentclass[manuscript]{aastex6}

\usepackage{graphicx}
\usepackage{epsfig}
\usepackage{times}
\usepackage{natbib}
\usepackage{url}
\usepackage{color}
\usepackage{amssymb,amsmath}
\usepackage{hyperref}
\shorttitle{Observations of Two EUV Waves and Their Mode Conversion}
\shortauthors{Chandra, Chen, Joshi, Joshi, \& Schmieder}

\begin{document}

\title{OBSERVATIONS OF TWO SUCCESSIVE EUV WAVES AND THEIR MODE CONVERSION}

\author{R. Chandra\altaffilmark{1}, P. F. Chen\altaffilmark{2,3}, R. Joshi\altaffilmark{1}, B. Joshi\altaffilmark{4}, and B. Schmieder\altaffilmark{5}}
\affil{$^1$ Department of Physics, DSB Campus, Kumaun University, Nainital -- 263 001, India; \email{rchandra.ntl@gmail.com}}
\affil{$^2$ School of Astronomy \& Space Science, Nanjing University, Nanjing 210023, China}
\affil{$^3$ Key Laboratory of Modern Astronomy \& Astrophysics (Nanjing University), Ministry of Education, Nanjing 210023, China}
\affil{$^4$ Udaipur Solar Observatory, Physical Research Laboratory, Udaipur 313 001, India}
\affil{$^5$ Observatoire de Paris, LESIA, UMR8109 (CNRS), F-92195 Meudon Principal Cedex, France}
\begin{abstract}
In this paper, we present the observations of two successive fast-mode 
extreme ultraviolet (EUV) wave events observed on 2016 July 23. Both fast-mode 
waves were observed by the Atmospheric Imaging Assembly (AIA) instrument on 
board the {\it Solar Dynamics Observatory} ({\it SDO}) satellite, with a 
traveling speed of $\approx$675 and 640 km s$^{-1}$, respectively. These 
two wave events were associated with two filament eruptions and two {\it GOES} 
M-class solar flares from the NOAA active region 12565, which was located near 
the western limb. The EUV waves mainly move toward the south direction. We 
observed the interaction of the EUV waves with a helmet streamer further away 
in the south. When either or one of the EUV waves penetrates into the helmet 
streamer, a slowly propagating wave with a traveling speed of $\approx$150 km 
s$^{-1}$ is observed along the streamer. We suggest that the slowly-moving waves 
are slow-mode waves, and interpret this phenomenon as the magnetohydrodynamic 
(MHD) wave mode conversion from the fast mode to the slow mode. Besides, we 
observed several stationary fronts in the north and south of the source region.
\end{abstract}
\section{INTRODUCTION}


Eruptive solar flares are associated with full or partial filament/prominence 
eruptions and now it is established that if the full or partial filament 
erupts, it will produce coronal mass  ejections (CMEs). Therefore, these three 
physical phenomena (filament eruptions, flares, and CMEs) are often coupled with 
each other and are the syndrome of the same process \citep{Forbes2000, 
Aulanier10, Chen11a}. Occasionally, these eruptions are accompanied by EUV 
waves, which can be observed over almost the entire solar surface. Historically 
these were known as ``EIT waves" \citep{Moses97,Thompson98} because of their 
discovery by EUV Imaging Telescope \citep[EIT,][]{delab1995} onboard the {\it 
Solar and Heliospheric Observatory} satellite \citep[{\it SOHO},][]{Domingo95}. 
Later on, different authors proposed different terminologies, such as ``coronal 
waves" \citep{Wang2000, Wu2001}, ``large-scale coronal propagating fronts" 
\citep{Nitta13}, and ``coronal propagating front" \citep{schr11}.

Since the discovery of EUV waves in 1998, extensive studies have been conducted. 
Besides the EIT telescope, EUV waves were also observed by other space-borne 
missions such as {\it Hinode}, {\it STEREO}, and, since 2010, {\it Solar 
Dynamics Observatory} \citep[{\it SDO},][]{Biesecker02, Chen02, Harr03, 
Okamoto04, Chen11, Delannee14, Chandra16}. Diverse observational features have 
been reported. Some of them remind us of fast-mode waves, e.g., the propagation 
speeds up to above 1000 km s$^{-1}$ \citep{Nitta13} and the wave reflection 
\citep{Gopalswamy09}. However, some of the features tend to oppose the fast-mode 
wave nature for the EUV waves, e.g., the existence of stationary EUV waves 
\citep{Delannee00}, the subsonic wave speed \citep{Trip07, Thom09, Zhukov09}, 
and the helicity-dependent rotation direction \citep{Podl05, attr07}. In order 
to reconcile these contradicting observational features, \citet{Chen02, Chen05} 
proposed that a filament eruption is accompanied by two types of EUV waves with 
different nature, i.e., the faster one corresponds to the fast-mode wave or 
shock wave \citep{Thompson98, Wang2000, Wu2001, Chandra18} and is the coronal 
counterpart of H$\alpha$ Moreton wave \citep{Moreton60}, whereas the slower EUV 
wave is an apparent motion that is produced by the successive stretching of the 
closed magnetic field lines overlying the erupting filament. Their magnetic 
field line stretching model for the slower EUV waves can naturally explain why 
their velocities are roughly one-third of the fast-mode wave speed in the solar 
corona. Several other models have been proposed to explain the slower EUV waves 
\citep[see reviews by][for details]{Warmuth15, Chen16, Long17, Krause18}, e.g., 
slow-mode waves \citep{will07, wang09}, successive reconnection model 
\citep{attr07}, and Joule heating at the interface between the erupting magnetic 
field and the background field \citep{Delannee07}. The co-existence of two EUV 
waves has been confirmed by various authors \citep{Chen11, schr11, Asai12, 
cheng12, kuma13, shen13, whit13}. In this scenario, it would be confusing 
to call any wave pattern in EUV images as an EUV wave. In order to avoid the 
ambiguity regarding the two types of EUV waves, \citet{Chen16} proposed to use 
different terminologies for them, e.g., coronal Moreton waves for the faster EUV 
waves and ``EIT waves" for the slower EUV waves. Note that in some events only 
one of the two types of EUV waves is clearly visible. 

Besides the low speed, i.e., $\sim$10--300 km s$^{-1}$, ``EIT waves" possess 
another peculiar characteristic, i.e., they stop at the footprint of magnetic 
quasi-separatrix layers (QSLs), as illustrated by Figure 7 in 
\citet{Delannee00}.  Such a feature was successfully explained by the magnetic 
field line stretching model \citep{Chen05} since the magnetic field outside the 
QSL belongs to another magnetic system, and cannot be pushed to stretch up by 
the erupting filament in the source region. Interestingly, when a fast-mode 
EUV wave passes a magnetic QSL, a bright stationary front is generated behind 
the continuously propagating but significantly weakened fast-mode wave 
\citep{Chandra16}. With 2-dimensional magnetohydrodynamic (MHD) simulations, 
\citet{Chen16a} proposed that just before the QSL there happens mode 
conversion, where the incident fast-mode wave is partly converted to a 
slow-mode wave. The slow-mode wave propagates along the closed magnetic loop. 
Seen from the top, the slow-mode wave trapped inside the magnetic loop looks 
like a stationary front since it cannot cross the quasi-vertical field lines. 
A stationary wave front is reproduced as a fast-mode wave interacts with the 
boundary of a coronal hole, which is a special QSL \citep{pian17}. Note that 
the observations in \citet{Chandra16} could not show the propagation of the 
slow-mode wave along closed magnetic loops due to the non-favoring viewing 
angle. Therefore, mode-conversion as the mechanism for the formation of the 
stationary front was a conjecture. The propagation of the slow-mode wave can be 
best revealed above the solar limb. Only recently, \citet{Zong17} clearly 
showed that after a fast-mode wave interacts with a helmet streamer with an 
incident speed of 380 km s$^{-1}$, the wave speed is reduced  significantly to 
160 km s$^{-1}$. Whereas their observations strongly support the wave mode 
conversion mechanism, the exact speed of the slow-mode wave can be estimated 
more accurately when the slice is taken along the travelling direction.

In this article, we present the observations of two successive filament 
eruptions from NOAA active region (AR) 12565 on 2016 July 23. The filament 
eruptions were associated with two medium-class solar flares and two fast-mode 
EUV waves. Both EUV waves interacted with a coronal streamer far away in the 
southern hemisphere, and the newly generated waves after the interactions are 
detected by {\it SDO}. This paper is organized as follows: Section \ref{obs} 
describes the data sets. The observational results are presented in Section 
\ref{res}, which is discussed in Section \ref{Discussion}.

\section{OBSERVATIONS}
\label{obs}

The events on 2016 July 23 were well observed by the Atmospheric Imaging 
Assembly \citep[AIA,][]{Lemen12} onboard the {\it Solar Dynamics Observatory} 
\citep[{\it SDO,}][]{Pesnell12} satellite. The high cadence (12 s) and high 
spatial resolution ($0\farcs6$) of the AIA telescope provide us an excellent 
opportunity to analyze the kinematics and dynamics of these events.  For our 
study, we use the AIA data observed in 304 \AA, 193 \AA\, and 171 \AA. In order 
to obtain a better quality of EUV waves, we first average four images and then 
make the running difference images in 193 \AA. The reason for choosing 193 \AA\ 
is that EUV waves are best discernible at this wavelength. All the images are 
co-aligned and corrected for the solar rotation using the routines available in 
the solar software \citep[SSWIDL,][]{Freeland98}. 

The flares were also observed by the {\it Reuven Ramaty High Energy Solar 
Spectroscopic Imager} \citep[{\it RHESSI},][]{Lin02} telescope at different 
energy channels in X-ray. We construct the X-ray images at different energy 
bands using the CLEAN algorithm, which provides us the spatial resolution of 
2$\arcsec$. For the associated CMEs, we use the data from LASCO coronagraph 
available at the CDAW website \citep{Gopalswamy09a}.

\section{RESULTS}\label{res}

\subsection{Filament eruptions and flares}
\label{filament}

\begin{figure*}
\centering
\includegraphics[width=0.8\textwidth]{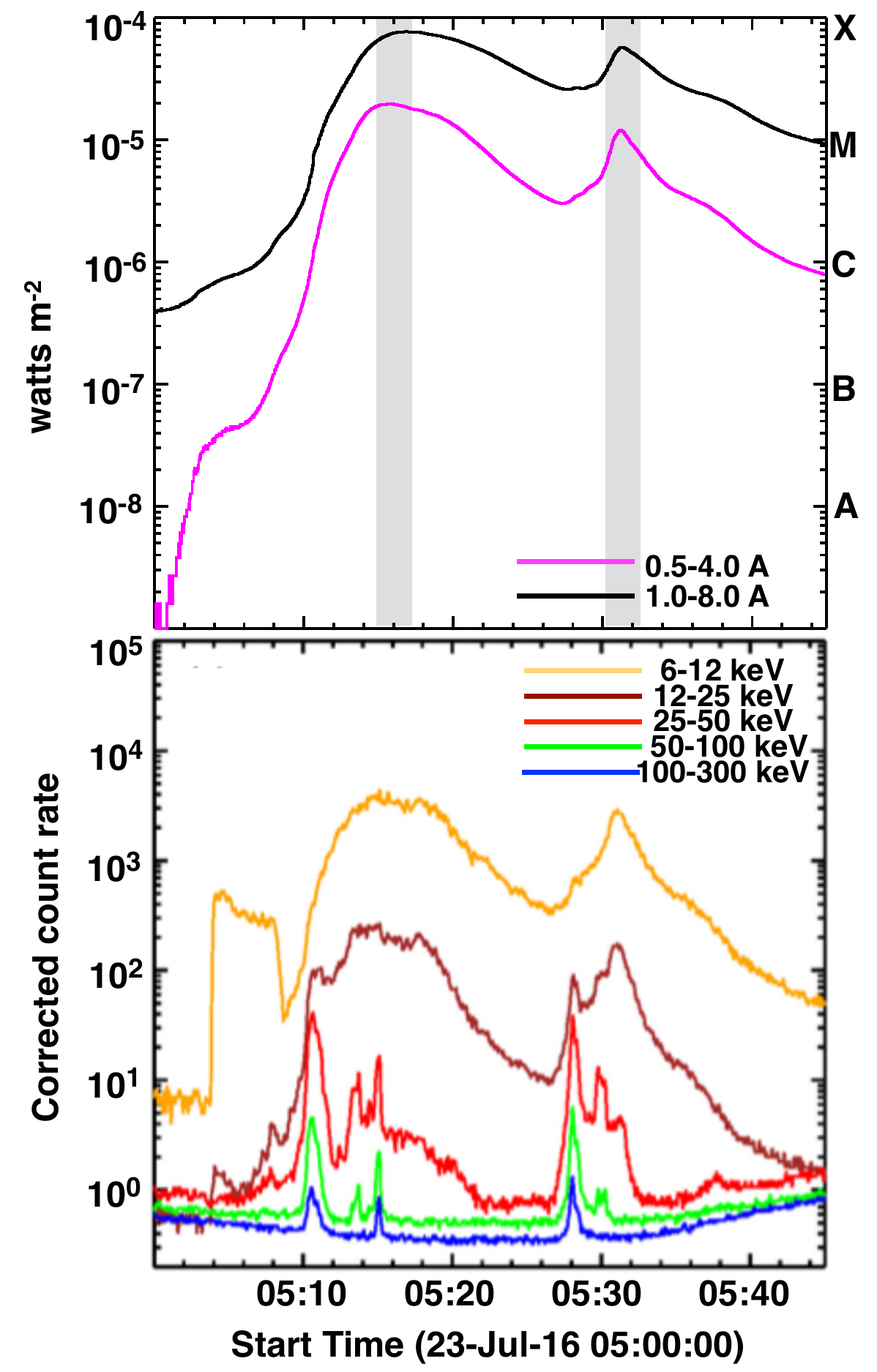}
\caption{Top panel: GOES evolution of flares in X-rays. The vertical bars indicate the 
peak time of flares. Bottom panel: Temporal evolution of flares  observed by the RHESSI satellite at 
different energy channels.}
\label{goes}
\end{figure*}

\begin{figure*}
\centering
\includegraphics[width=1.0\textwidth]{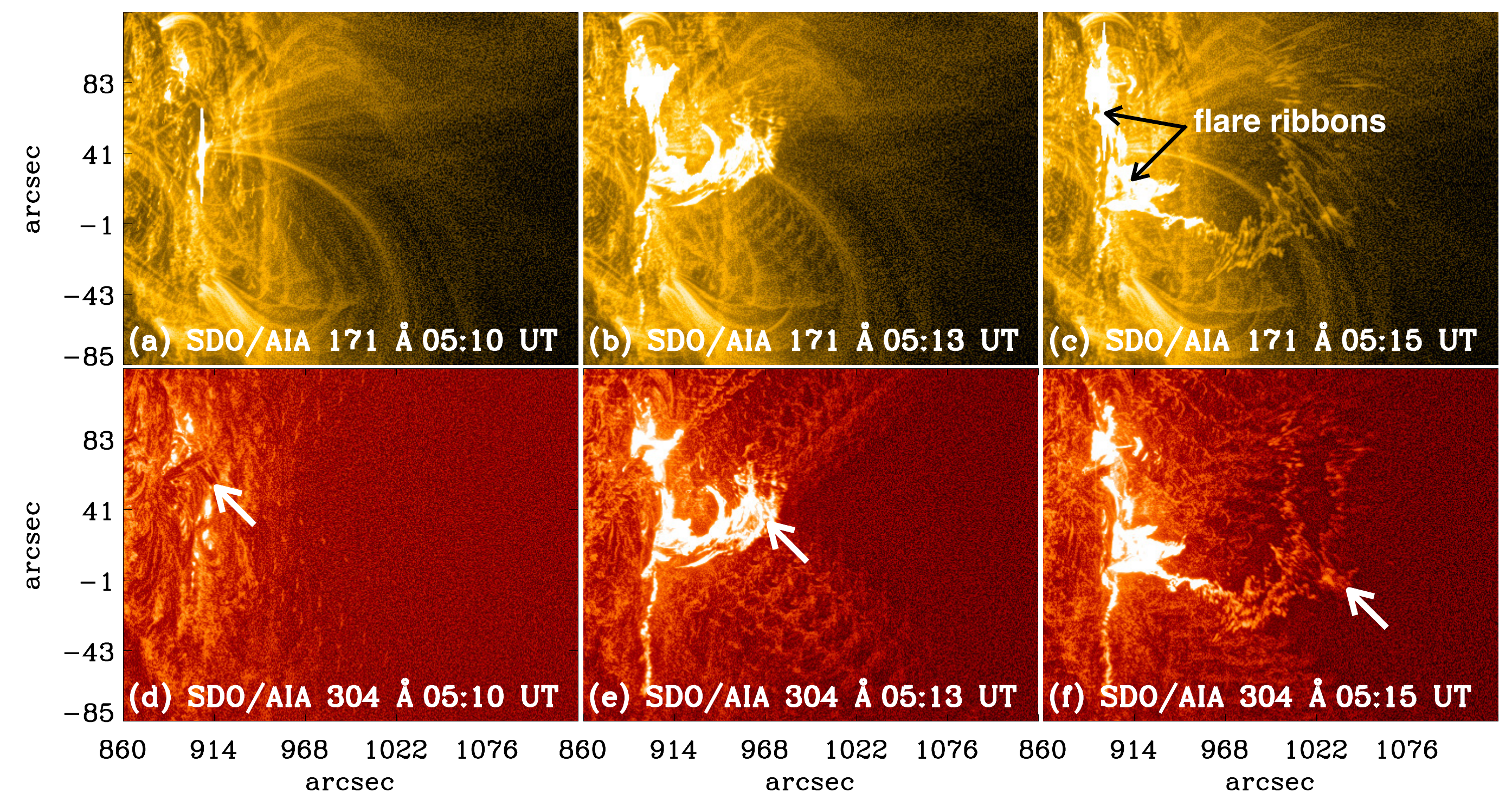}
\caption{Evolution of the filament eruption and the first flare in AIA 171 \AA\ 
(top )and 304 \AA\ (bottom). The Filament eruption is shown by the white 
arrows. Two ribbons of the flare are shown by the black arrows.}
\label{flare1}
\end{figure*}

\begin{figure*}
\centering
\includegraphics[width=1.0\textwidth]{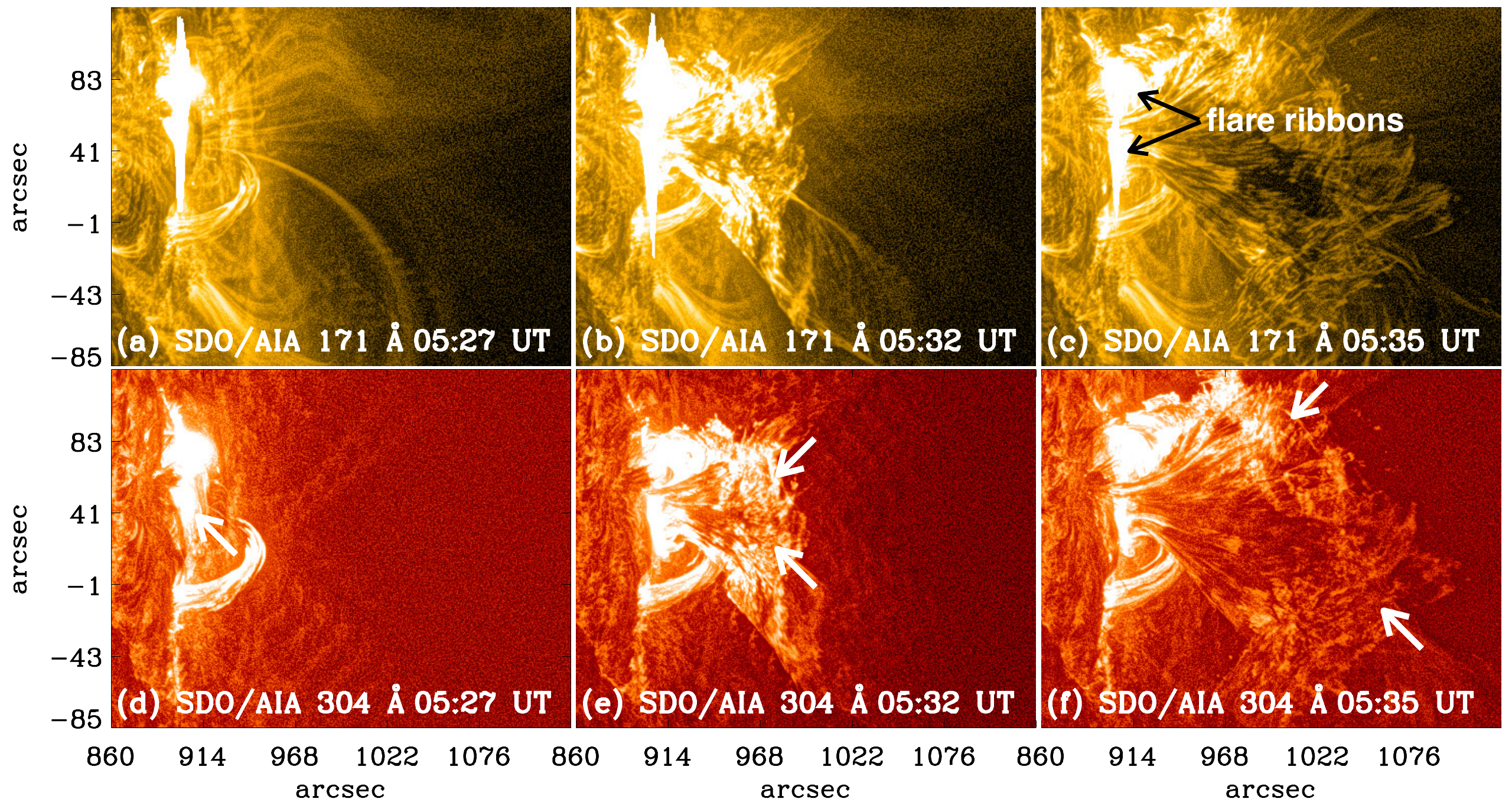}
\caption{Evolution of the second filament eruption and the second flare in AIA 
171 \AA\ (top) and 304 \AA\ (bottom). The erupting filament is indicated by the white arrows.
Two ribbons of the flare are shown by the black arrows.}
\label{flare2}
\end{figure*}

NOAA AR 12565 was located near the western limb (N04W89) on 2016 July 23. The 
active region with a  multipolar (\url{https://www.solarmonitor.org}) 
magnetic distribution produced three {\it GOES} M-class and four C-class flares 
on that day. The evolution of both the flares, observed by {\it GOES} in 
X-rays is given in Figure \ref{goes} (top panel). Among them, two flares 
(hereafter referred as the first flare and the second flare) were associated 
with filament eruptions. The first filament eruption occurred at $\sim$05:08 UT. 
The evolutions of the first filament eruption and the associated {\it GOES} 
M7.6-class flare in AIA 171 \AA\ and 304 \AA\ are presented in Figure 
\ref{flare1}. According to the {\it GOES} observations, the onset and the peak 
times of the first flare were at 05:00 UT and 05:16 UT, respectively. Due to the 
location of flare site, which was very close to the limb, the two ribbons of the 
flare were very close to each other, as seen in Figure \ref{flare1} and 
\ref{flare2}(c). Nevertheless, based on their relative locations, we can still 
judge that the magnetic system has a negative helicity, which follows the 
hemispheric rule \citep{ouy17}.

Before the end of the first flare, at $\sim$05:25 UT we observed the second 
filament eruption from the same active region, which produced the second {\it 
GOES} M5.5-class flare. The onset and peak times of the second flare were 
05:27 UT and 05:31 UT, respectively. The second flare was a long duration event 
and it continued until 07:00 UT. The evolutions of the second filament eruption 
and flare in AIA 171 \AA\ and 304 \AA\ are presented in Figure \ref{flare2}. It 
shows that the filament eruption was experiencing a whipping motion, one typical 
asymmetric eruption of solar filaments \citep{liu09}. Each of the two filament 
eruptions was associated with a CME observed by the LASCO coronagraph. The 
linear speed of the CME was $\sim$835 km s$^{-1}$ with an acceleration of 
$-15.2$ m s$^{-2}$, according to the CDAW website (\url{https://cdaw.gsfc.nasa.gov}).

\begin{figure*}
\centering
\includegraphics[width=0.7\textwidth]{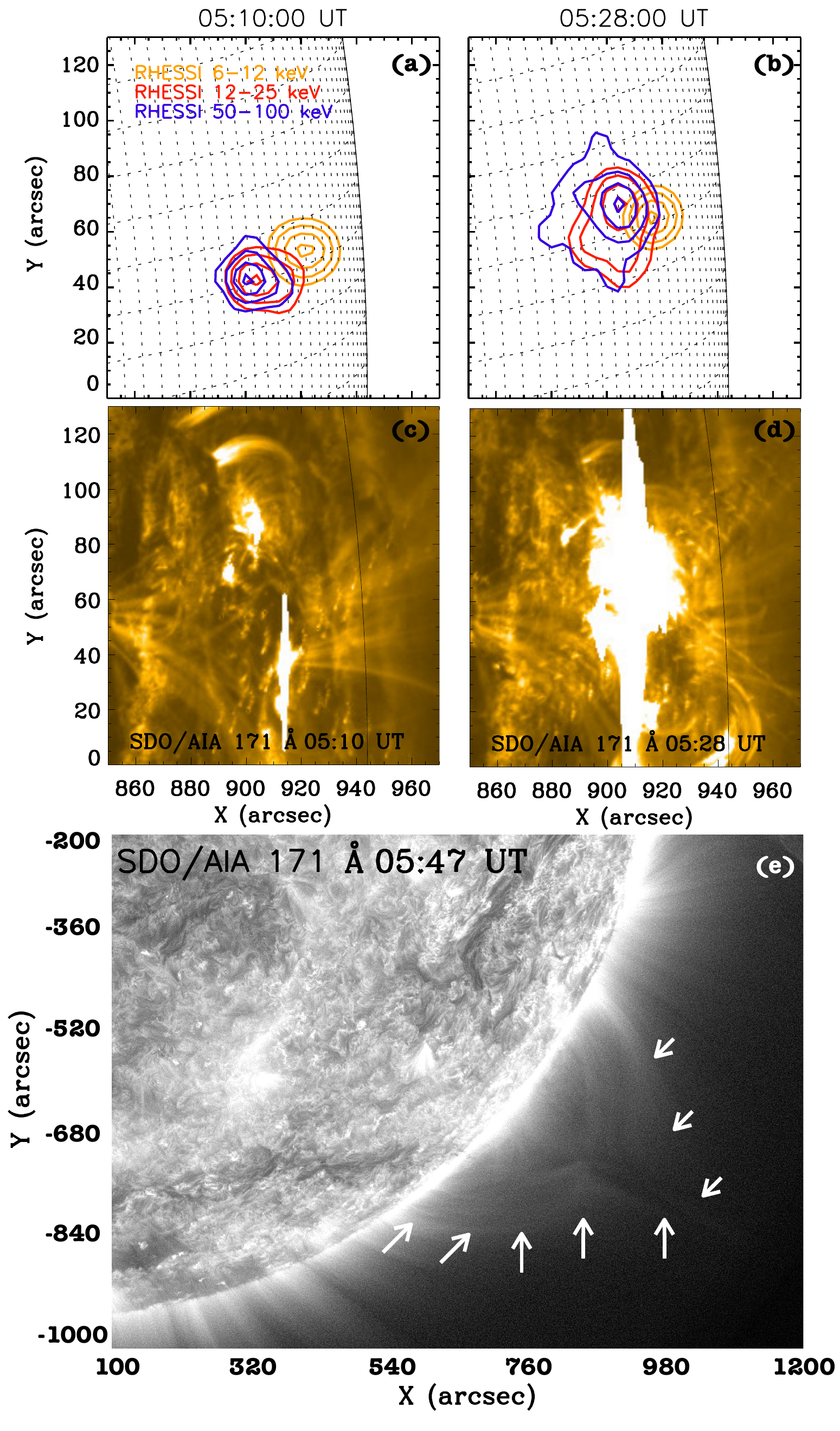}
\caption{Upper row: {\it RHESSI} contours of the first (panel a) and the second (panel b) 
flare in different energy bands. Middle row: Corresponding AIA 171 \AA\ images 
of the first (panel c) and the second flare (panel d). Bottom row: SDO/AIA 171 \AA\ image showing 
the location (by white arrows) of helmet streamer with which the EUV waves interacted.}
\label{rhessi}
\end{figure*}

The strength of energy release process driven by magnetic reconnections in the 
wake of filament eruptions can be ascertained from the RHESSI X-ray sources 
formed underneath the erupting flux ropes (Figure \ref{rhessi}(a)-(b)). The 
RHESSI images show two components of the flare emission: low energy emission
imaged at 6-12 keV energy band located closer to the limb and high energy 
emission as demonstrated by the 12-25 keV and 50-100 keV sources that are 
almost co-spatial. Notably, both flare were associated with very strong HXR 
emissions up to $\sim$300 keV.
The temporal evolution of both the flares observed by RHESSI in different
energy channels is displayed in Figure \ref{goes} (bottom panel).

\subsection{Kinematics of the EUV Waves}
\label{EUV}

\begin{figure*}
\centering
\includegraphics[width=0.9\textwidth]{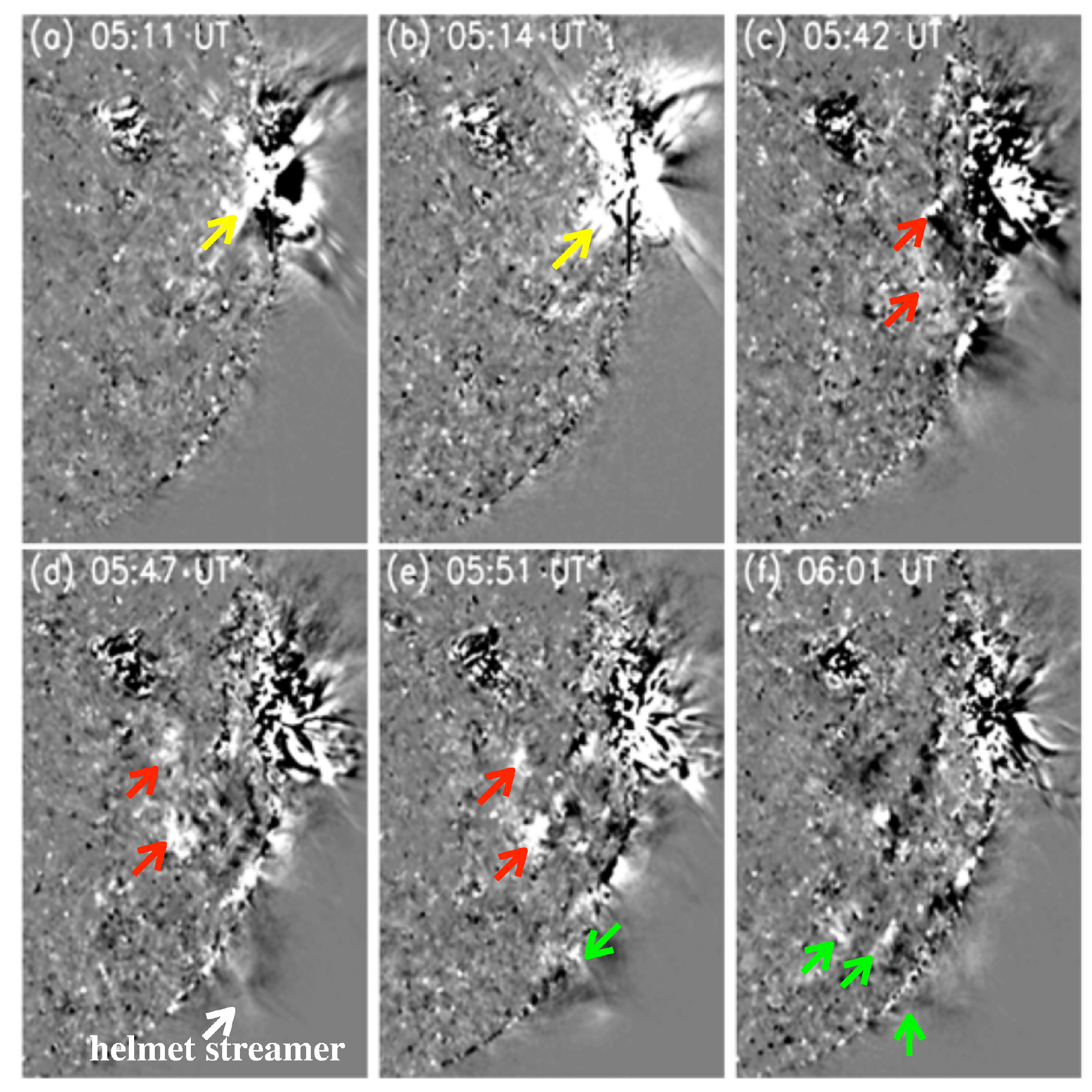}
\caption{Evolution of EUV waves in AIA 193 \AA\ difference images. The yellow 
and the red arrows indicate the first and the second EUV wave fronts. The green 
arrows indicate the EUV waves after passing through helmet streamer. The tip of
the helmet streamer is indicated by the white arrow in panel (d).}
\label{helmet}
\end{figure*}

\begin{figure*}
\centering
\includegraphics[width=0.8\textwidth]{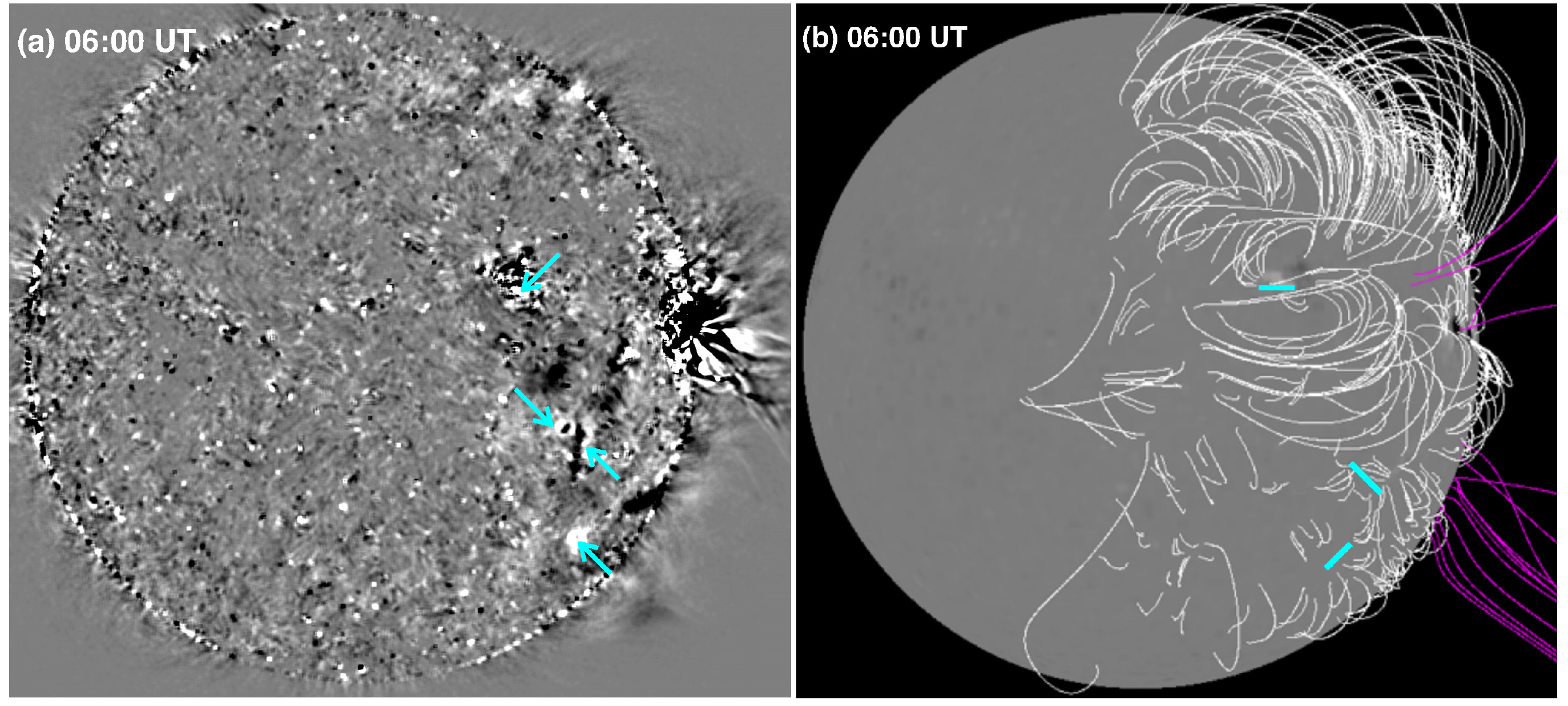}
\caption{AIA 193 \AA\ image indicating the locations of stationary fronts with 
cyan arrows (left) and the PFSS extrapolation of magnetic field (right). 
The stationary fronts  are located in the QSLs sites and 
shown by cyan lines.}
\label{pfss}
\end{figure*}

The two eruption events on 2016 July 23 were both associated with EUV 
waves. The first EUV wave appeared at $\sim$05:10 UT and it propagated along 
the southeast direction. Panels (a--b) of Figure \ref{helmet} display the 193 
\AA\ base difference images during the first eruption, where the EUV wave 
fronts are marked by the yellow arrows. However, it is noted that this wave is 
very weak and poorly visible in the AIA difference images. Around 20 minutes 
later, the second filament eruption occurred, producing another EUV wave. This 
second EUV wave started around 05:30 UT in AIA 193 \AA\ images. Its evolution 
in AIA 193 \AA\ is presented in the panels (c--e) of Figure \ref{helmet}, where 
the EUV wave fronts are indicated by the red arrows. The time difference 
between the initiations of the first and the second EUV waves is $\sim$20 
minutes, which is almost the same as the time difference between the first and 
the second filament eruptions. As shown by Figure \ref{helmet}(f), the second 
 EUV wave approached a helmet streamer above the southwestern limb (see also the 
attached movie of Figure \ref{helmet}). When these EUV waves propagated on 
the solar disk, several stationary fronts were observed, as marked by the cyan 
arrows in Figure \ref{pfss}(a)  (see also the AIA 193 \AA\ movie). To see the 
magnetic property of these stationary fronts, we extrapolate the photospheric 
magnetic field using the PFSS model available in SSWIDL. The result is shown in 
the right panel of Figure \ref{pfss}. We compare the locations of these 
stationary fronts with the extrapolated potential magnetic field and find that 
these locations are all co-spatial with QSLs. This confirms the earlier findings 
\citep{Delannee07,Chen02,Chandra16}.  Around the solar limb, we can see 
an EUV wave passing through a helmet streamer, as shown by the green arrows in 
Figure \ref{helmet}(e--f). Note that the location of the helmet streamer 
in AIA 171 \AA\ is also indicated in Figure \ref{rhessi}(e). 

\begin{figure*}
\centering
\includegraphics[width=1.0\textwidth]{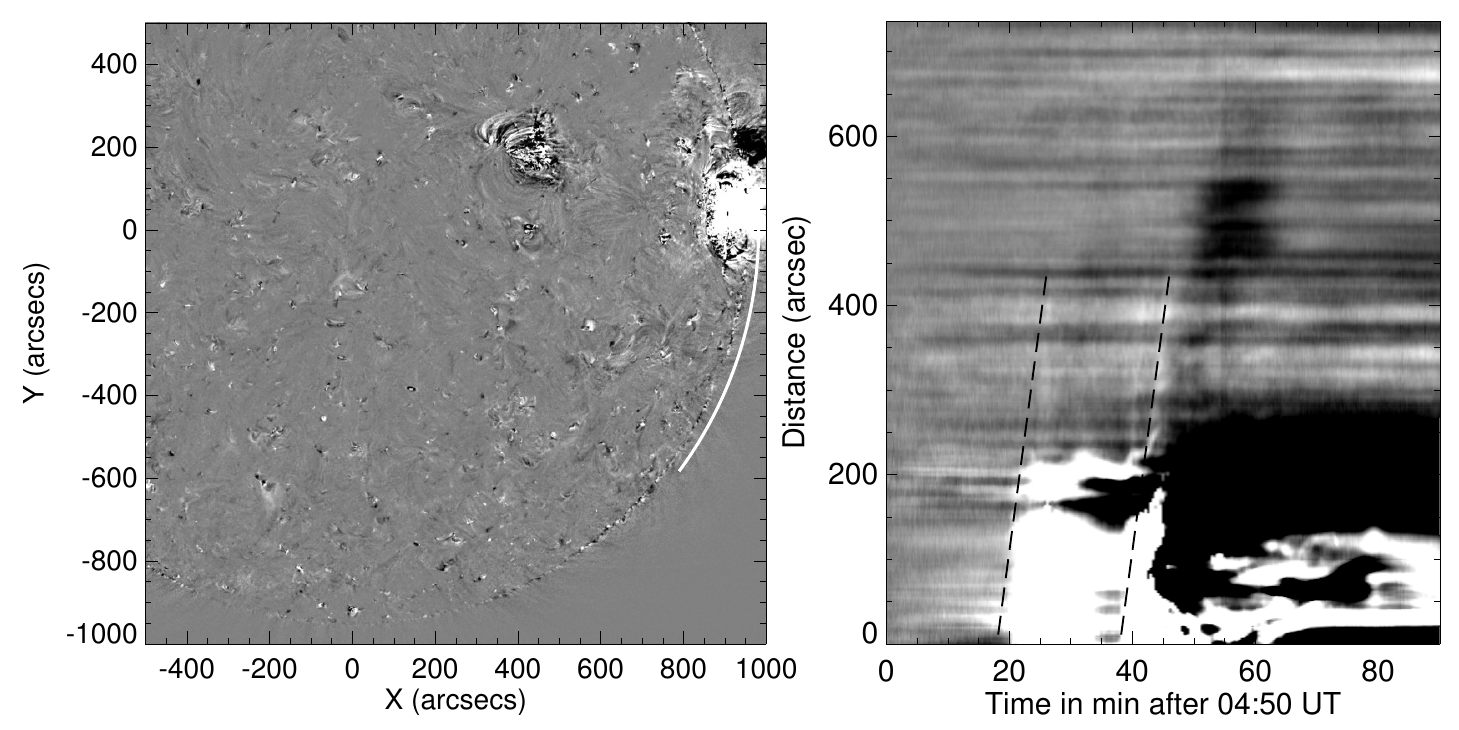}
\caption{Left: The AIA 193 \AA\ difference image at 05:12 UT, where the white circular
curve along the solar limb indicates the slice used for the time-distance 
diagram analysis in the right panel. Right: Time-distance diagram showing 
two fast EUV waves 
associated with the two filament eruptions. The speed of first wave is 
$\approx$675 km s$^{-1}$ and that of the second wave is $\approx$640 
km s$^{-1}$.}
\label{slice1}
\end{figure*}

\begin{figure*}
\centering
\includegraphics[width=0.95\textwidth]{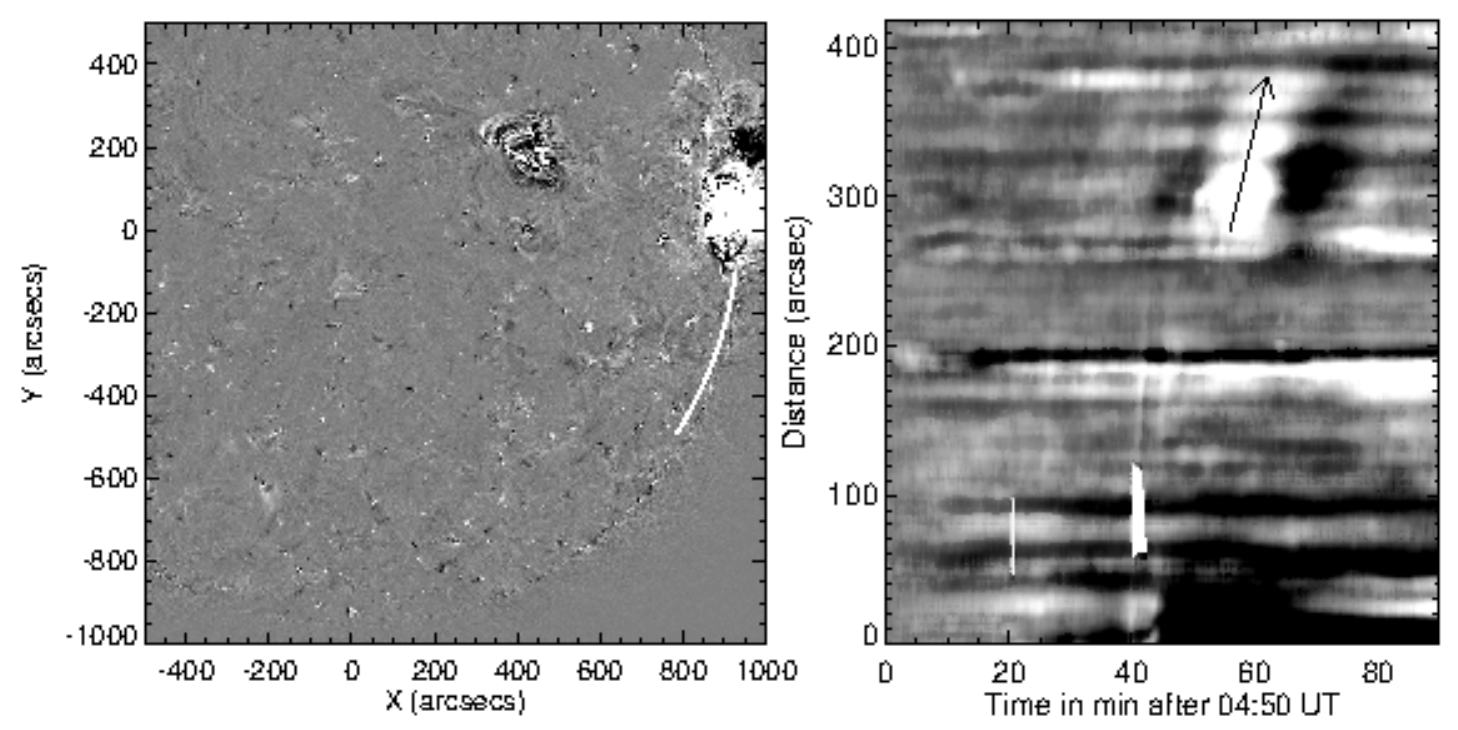}
\caption{ Left: The AIA 193 \AA\ difference image at 05:12 UT,
the white circular slice in the solar surface used for the time-distance plot 
presented in the right panel. Right: Time-distance plot showing the two fast EUV waves together with slower EUV 
wave (EIT wave) component (indicated by black arrow) 
associated with second event (speed $\approx$178 km s$^{-1}$).}
\label{slice0}
\end{figure*}

To show the early kinematics of the EUV waves, we select a circular slice 
parallel to the solar limb as indicated by the white curve in the left panel of 
Figure \ref{slice1}. The corresponding time-slice diagram is displayed in the 
right panel. It is seen that two EUV waves emanate from the active region at 
$\sim$05:10 UT and 05:30 UT, with an initial velocity of 675 and 640 km 
s$^{-1}$ respectively, as indicated by the black dashed lines. The initiation 
times are consistent with the impulsive phases of the two flares. Both EUV 
waves decelerate at a distance of 500\arcsec\ from the starting point of the 
slice. It is noticed that each EUV wave has fine structures with several 
strips. The propagation velocity of these two EUV waves are in the typical 
range of fast-mode waves in the solar corona \citep{Chen16}, therefore both of 
them should be fast-mode EUV waves. In order to verify that both waves 
above the limb are fast-mode waves, rather than the CME front, we select a 
parallel slice but inside the solar disk, as indicated by the left panel of 
Figure \ref{slice0}. The time-distance diagram along this slice is displayed 
in the right panel of Figure \ref{slice0}. It is seen that in addition to two 
fast-moving EUV waves similar to those in Figure \ref{slice1}, a brighter EUV 
front propagated away with a speed of 178 km s$^{-1}$. This slower wave has a 
traveling velocity roughly 3 times smaller than those of the fast-mode waves, 
as predicted by the magnetic fieldline stretching model \citep{Chen02, Chen05}. 
It is believed that such a non-wave component of the EUV waves corresponds to 
the CME frontal loop \citep{Chen09}. 

\begin{figure*}
\centering
\includegraphics[width=0.90\textwidth]{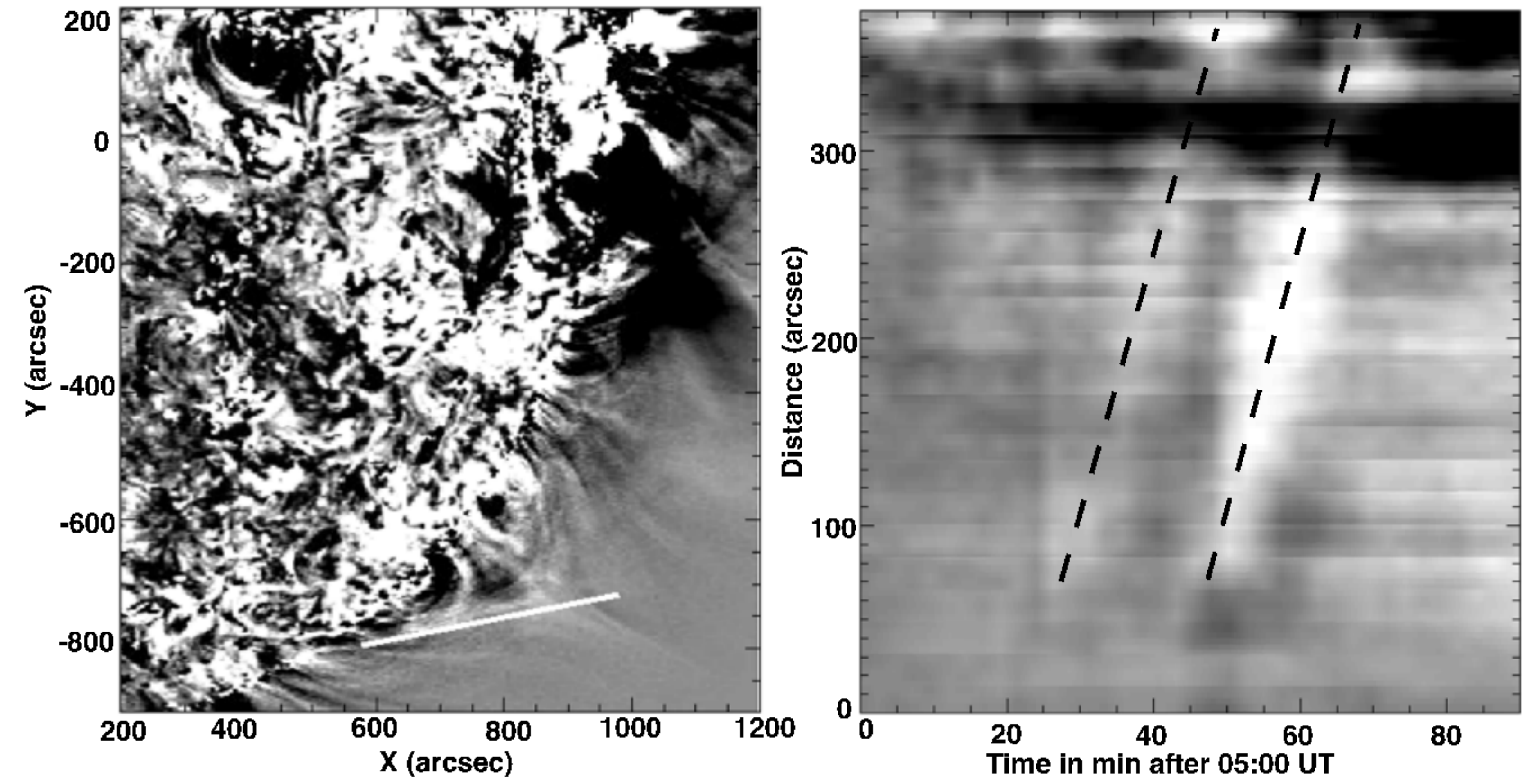}
\caption{Left: The AIA 193 \AA\ difference image at 05:59 UT, where the white
line along one leg of a helmet streamer indicates the slice used for the
time-distance diagram analysis. Right: Time-distance diagram showing two slowly
moving EUV waves along the leg of the helmet streamer  with a speed of $\approx$150 km s$^{-1}$.}
\label{slice2}
\end{figure*}

Whereas the second EUV wave was seen clearly to propagate to the southern 
hemisphere as revealed by Figure \ref{helmet}, the first EUV wave faded rapidly, 
becoming very faint outside the source active region. However, when both EUV 
waves hit the helmet streamer above the southwestern limb, they become slightly 
brighter. To see the kinematics of the EUV waves across the helmet streamer, we 
select a slice shown as the white line in the left panel of Figure \ref{slice2}. 
Such a slice is along one leg of the streamer. The corresponding time-distance 
diagram along this slice is displayed in the right panel of Figure \ref{slice2}. 
Again we can see clearly two waves starting from $\sim$05:25 UT and 05:45 
UT. Similar to Figure \ref{slice1}, the two waves in Figure \ref{slice2} are 
also separated by $\sim$20 minutes, which is indicative of that there is 
one-to-one correspondence between the wave pairs in Figures \ref{slice1} and 
\ref{slice2}. However, the traveling speeds of the wave pair in Figure 
\ref{slice2} are around 150 km s$^{-1}$, which is much slower than that of the 
wave pair in Figure \ref{slice1}.

\section{DISCUSSION}
\label{Discussion}

Since the discovery of EUV waves in late 1990s, they were initially treated as 
fast-mode MHD waves. The high speed of the waves in some events tend to 
favor the fast-mode wave model. However, the fast-mode wave model cannot 
explain the subsonic EUV waves whose velocity is as low as $\sim$10 km s$^{-1}$ 
\citep{Zhukov09}, and the model was severely challenged by the discovery of 
stationary wave fronts \citep{Delannee99, Delannee00}. As claimed by 
\citet{Chen02, Chen05}, a possible solution to these discrepancies is that 
there are two types of EUV waves, i.e., a fast-mode wave or shock wave that is 
nearly cospatial with H$\alpha$ Moreton wave \citep{vrsn02, chend05, fran16}, 
and an apparent wave whose velocity is typically around one-third of the 
fast-mode wave. Such co-existence of two EUV waves have been confirmed by 
various authors \citep{Chen11, schr11, Asai12, cheng12, kuma13, shen13, whit13}.

Being retrospected, the discovery of stationary EUV wave front played a very 
important role in deepening our understanding of coronal EUV waves. However, 
it was revealed by \citet{Chandra16} that a stationary front can be generated 
by the interaction between a fast-mode wave and a magnetic QSL. Inspired by 
their observations, \citet{Chen16a} proposed that, just prior to the magnetic 
QSL, there exists a local layer where the Alfv\'en speed is equal to the sound 
speed, where MHD waves can be converted from one mode to the other 
\citep{Cally05}, in this case, from the fast-mode wave to slow-mode wave. Since 
the converted slow-mode wave propagates along the closed magnetic loops, and 
cannot run across the nearly vertical magnetic field lines, it looks like a 
stationary front when seen from above. The limb event studied by \citet{Zong17} 
provided the first support for the mode-conversion model. 

In the compound eruption events of 2016 July 23, two filaments erupted 
successively within $\sim$20 minutes like sympathetic events. The two events 
were associated with two {\it GOES} medium-sized flares, two fast-mode EUV 
waves, and two CMEs. In our observations the non-wave component of the EUV 
waves, or ``EIT wave", was clearly visible in the second event. Around the 
source region, the two fast-mode waves propagated outward with a speed of 
675 and 640 km s$^{-1}$, respectively. However, when they 
penetrated into a distant helmet streamer, the waves became slightly brighter 
and much slower, with a speed of 150 km s$^{-1}$ along the leg of the streamer. 
Such a speed is a typical value of sound speed for the coronal plasma with a 
temperature of 1 MK. The two slow-mode waves were separated by 20 minutes, the 
same as the time delay between the initial two fast-mode waves, which is 
strongly indicative of that each fast-mode wave is converted to a slow-mode 
wave.

It is understandable why the wave mode conversion happens inside a helmet 
streamer in both \citet{Zong17} and ours: Whereas the Alfv\'en speed is much 
larger than the sound speed near the base of a streamer in the low corona, the 
magnetic field at the tip of a streamer is close to zero, where the Alfv\'en 
speed is close to zero, there should exist a layer slightly below the streamer 
tip where the Alfv\'en speed is equal to the sound speed. It is in such a place 
where fast-mode waves are converted into slow-mode waves. Since the streamer 
tip maps to a magnetic QSL, the trapped slow-mode wave cannot cross the field 
lines, and would be seen to stop near the QSL when observed from the top.
It is noted in passing that such a wave-streamer interaction is consistent with the 
picture that the fast-mode waves driven by filament eruptions have a large-scale dome-shaped 
structure in the corona \citep{Chen02}.

To summarize, we analyzed the wave phenomena associated with the possible 
sympathetic eruption events on 2016 July 23. The results include: 
(1) The two episodes of filament eruptions from the active region NOAA 12565, 
which were separated by 20 minutes, drove two fast-mode EUV waves or shock 
waves with a speed of 675 and 640 km s$^{-1}$; (2) The two fast-mode EUV waves 
interacted with a helmet streamer in another hemisphere and were observed to 
propagate along the leg of the streamer, with a speed of 150 km s$^{-1}$. We 
claim that slowly propagating waves are slow-mode MHD waves which are converted 
from the fast-mode waves. These observations strongly support the wave-mode 
conversion model proposed by \citet{Chen16a} in order to explain some 
stationary EUV wave fronts at magnetic QSLs.

\acknowledgments
We are very grateful to the referee for the useful comments and suggestions.
The authors thank the open data policy of SDO. RC and BJ acknowledge
the support from SERB-DST project no. SERB/F/7455/ 2017-17. RJ thanks 
the Department of Science and Technology (DST), Government of India for an 
INSPIRE fellowship. PFC was supported by the Chinese foundations (NSFC 
11533005, U1731241, and Jiangsu 333 Project No. BRA2017359). PFC thanks ISSI-BJ for 
supporting wave meetings where some of the ideas in this paper were stimulated.


\clearpage

\end{document}